\documentclass[%
 preprint,%
 amssymb, amsmath,%
 aip,jcp,%
]{revtex4-2}

\usepackage[T1]{fontenc}
\usepackage{enumitem}  
\usepackage{setspace}

\usepackage{tabularx}
\usepackage{multirow}
\usepackage{booktabs}
\usepackage{siunitx}

\usepackage[colorlinks=true,linkcolor=blue]{hyperref}%
\expandafter\ifx\csname package@font\endcsname\relax\else
 \expandafter\expandafter
 \expandafter\usepackage
 \expandafter\expandafter
 \expandafter{\csname package@font\endcsname}%
\fi
\usepackage{graphicx}
\usepackage{color}
\usepackage[version=4]{mhchem}
\usepackage{verbatim}
\usepackage{subcaption}    

\usepackage{amsmath}
\usepackage{lineno}
\usepackage{geometry}
\usepackage{iftex}
\usepackage{color,soul} 
\usepackage[version=4]{mhchem}

\newcommand{\eV}{\ensuremath{\text{eV}}}

\newcommand{\ie}{i.\,e.}
\newcommand{\eg}{e.\,g.}

\newcommand{\evU}{\ensuremath{\text{eV}}}
\newcommand{\rateU}{\ensuremath{\text{cm}^{3}\,\text{s}^{-1}}}

\newcommand{\CNm}{\ce{CN-}}
\newcommand{\CNp}{\ce{CN+}}
\newcommand{\HCNp}{\ce{HCN+}}
\newcommand{\HCN}{\ce{HCN}}
\newcommand{\HNC}{\ce{HNC}}
\newcommand{\HNCp}{\ce{HNC+}}
\newcommand{\HCNHp}{\ce{HCNH+}}
\newcommand{\HNCFp}{\ce{HNCF+}}

\begin{document}

\title{Measurements of rate coefficients of \CNp, \HCNp\ and \HNCp\ collisions with \ce{H2} at cryogenic temperatures}%

\author{Petr Dohnal}
\affiliation{Department of Surface and Plasma Science, Faculty of Mathematics and Physics, 
Charles University, Prague, V Hole{\v s}ovi{\v c}k{\' a}ch 2, 180 00, Czech Republic}
\author{Pavol Jusko}
\email[]{pjusko@mpe.mpg.de}
\affiliation{Max Planck Institute for Extraterrestrial Physics, Gie{\ss}enbachstra{\ss}e 1, 85748 Garching, Germany}
\author{Miguel Jim{\'e}nez-Redondo}
\affiliation{Max Planck Institute for Extraterrestrial Physics, Gie{\ss}enbachstra{\ss}e 1, 85748 Garching, Germany}
\author{Paola Caselli}
\affiliation{Max Planck Institute for Extraterrestrial Physics, Gie{\ss}enbachstra{\ss}e 1, 85748 Garching, Germany}


\date{\today}

\begin{abstract}
The experimental determination of the reaction rate coefficients for production and destruction of \HCNp\ and \HNCp\ in collisions 
with \ce{H2} is presented. A variable temperature 22 pole radio frequency ion trap was used to study the reactions in the 
temperature range of 17 -- 250 K. The obtained rate coefficients for the reaction of \CNp\ and of \HCNp\ with 
\ce{H2} are close to the collisional (Langevin) value, whereas that for the reaction of \HNCp\ with \ce{H2} is quickly decreasing 
with increasing temperature. The product branching ratios for the reaction of \CNp\ with \ce{H2} are also reported
and show a notable decrease of \HNCp\ product with respect to \HCNp\ product with increasing temperature.
These measurements have consequences for current astrochemical models of cyanide chemistry, in particular for the \HCNHp\ cation.
\end{abstract}


\maketitle 

\section{Introduction}

HCN and its higher energy isomer HNC ($0.64\;\eV$) are presumably the two simplest isomers in chemistry.
On earth, at $300\;\text{K}$, the \HCN\ isomer is populated almost exclusively.
Both isomers have been detected in a variety of environments of the interstellar medium (ISM) 
such as starless cores \cite{HilyBlant2010}, diffuse \cite{Liszt2001}, 
translucent \cite{Turner1997} and dense interstellar clouds \cite{Boger2005} 
and star \cite{Godard2010} and planet \cite{Graninger2015,Cleeves2018,Bergner2022} forming regions. 
Despite the $1.3\;\eV$ barrier for isomerization from HCN to HNC \cite{Loison2014},
HNC abundances are 
often found to be comparable to that of HCN, especially in cold environments with temperatures close
to $10\;\text{K}$ \cite{Hirota1998,Tennekes2006,Loison2014}. On the other hand, the HCN/HNC abundance ratio was
reported to be much greater than one in relatively warm objects such as hot cores or young stellar 
objects. For example, a HCN/HNC abundance ratio of 13 was observed for IRAS 16293–2422 \cite{Schoier2002}, 
while in the vicinity of the Orion-KL Nebula this ratio was approximately 80 \cite{Schilke1992}.
It has been suggested \cite{Loison2014} that the actual HCN/HNC abundance ratio is governed by competing 
processes in the given environment.
However, as noted by \citet{HernandezVera2017}, special care needs to be taken in interpreting the observations, 
as the rate coefficients for HCN and HNC differ significantly, especially when the collision partner is para-\ce{H2}, 
the dominant form of \ce{H2} in cold gas (e.g. \citet{Bruenken2014}).

One of the main sources of HCN and HNC molecules in the interstellar medium is the dissociative recombination 
of \ce{HCNH+} ions with electrons
\begin{subequations}
	\begin{align}
		\HCNHp + \ce{e-} &\to \ce{HCN} + \ce{H}         \label{eq:HCNHprecoma}   \\ 
		&\to \ce{HNC} + \ce{H} \label{eq:HCNHprecomb}  \\
		&\to \ce{CN} + \ce{H} + \ce{H} \label{eq:HCNHprecomc}  
	\end{align}
	\label{eq:HCNHprecom}
\end{subequations}
with almost the same probability for the three reaction channels \cite{Semaniak2001}. The principal 
processes leading to the formation of \ce{HCNH+} ions in molecular clouds are \cite{Loison2014,Quenard2017,Fontani2021}:
\begin{equation}
	\HCNp + \ce{H_2} \to \HCNHp + \ce{H},
	\label{eq:HCNpH2}
\end{equation}
\begin{equation}
	\HNCp + \ce{H_2} \to \HCNHp + \ce{H}, 
	\label{eq:HNCpH2}
\end{equation}

\begin{equation}
	\ce{C^+} + \ce{NH_3} \to \HCNHp + \ce{H}, 
	\label{eq:CpNH3}
\end{equation}
\begin{equation}
	\ce{H3+} + \HCN/\ce{HNC} \to \HCNHp + \ce{H2}, 
	\label{eq:HHHreact}
\end{equation}
\begin{equation}
	\ce{H_3O+} + \HCN/\ce{HNC} \to \HCNHp + \ce{H_2O},
	\label{eq:H3Opreact}
\end{equation}
\begin{equation}
	\ce{HCO+} + \HCN/\ce{HNC} \to \HCNHp + \ce{CO},
	\label{eq:HCOreact}
\end{equation}
where reactions (\ref{eq:HCNpH2}) and (\ref{eq:HNCpH2}) are thought to be the main pathway 
for \ce{HCNH+} formation in dense, cold regions such as L1544 \cite{Quenard2017} while 
reaction (\ref{eq:HCOreact}) shall dominate in warmer environments \cite{Fontani2021}. In early 
times of cloud formation \ce{HCNH+} ions are presumably produced mainly in reaction 
(\ref{eq:HHHreact}) \cite{Fontani2021}.

The processes leading to the production of \HNCp\ and its higher energy isomer \HCNp\ ions ($0.94\;\eV$) 
depend on the physical and chemical conditions of the molecular cloud.
In cold and dense regions, 
these ions are mainly produced by proton transfer reaction of \ce{H+} with HCN/HNC or in 
collisions of \ce{CN+} ions with \ce{H2} \cite{Quenard2017}
\begin{subequations}
	\begin{align}
		\CNp + \ce{H2} &\to \HNCp + \ce{H} \quad \Delta H = -2.28\;\eV  \label{eq:CNreactHNC}   \\ 
		&\to \HCNp + \ce{H} \quad \Delta H = -1.33\;\eV,  \label{eq:CNreactHCN}
	\end{align}
	\label{eq:CNreact}
\end{subequations}
with the enthalpies of formation, $\Delta H$, taken from ref. \cite{Scott1997}. Reaction (\ref{eq:CNreact}) 
was previously studied in SIFT (Selective Ion Flow Tube) experiments at 300~K by \citet{Petrie1991} 
and by \citet{Scott1997} who reported the value of the reaction rate coefficient to be 
$1.1\times10^{-9}~\rateU$ and $1.6\times10^{-9}~\rateU$, respectively. In both experiments, the 
observed branching ratio was 0.5 (\ie, the two reaction channels were equally probable). 
In another SIFT experiment, \citet{Raksit1984} obtained a value of 
the reaction rate coefficient of $1.0\times10^{-9}~\rateU$ at room temperature without distinguishing 
between reaction products. A slightly higher value of $1.24\times10^{-9}~\rateU$ was measured by 
\citet{McEwan1983} using the Ion Cyclotron Resonance technique (ICR) at near thermal energies. 
To the best of our knowledge there are no experimental data for reaction (\ref{eq:CNreact}) 
obtained at low, astrophysically relevant temperatures.

Reactions (\ref{eq:HCNpH2}) and (\ref{eq:HNCpH2}) were only studied by \citet{Petrie1991} 
at 300~K using the SIFT technique. The measured reaction rate coefficients were 
$8.6\times10^{-10}~\rateU$ and $7.0\times10^{-10}~\rateU$, respectively. The scarcity of experimental
data for these reactions is not surprising as it is very difficult to distinguish between the two 
isomer forms in experiments solely based on mass spectrometry.
Chemical probing is an obvious method to enhance mass sensitive experiments to
gain isomer sensitivity \cite{Smith2002}.

Photon processes also play an important role in the \ce{HCN}/\ce{HNC} (neutral/cation) abundance \cite{Aguado2017}.
Although \CNp\ \cite{Thorwirth2019} and \HCNHp\ \cite{Amano2006} have been extensively studied experimentally in 
the mm-wave range, only a Ne matrix IR spectrum \cite{Forney1992} 
and  vibrational bands determined by neutral photoelectron spectroscopy \cite{Fridh1975,Eland1998,Gans2019} 
are available for \HNCp/\HCNp\ cations.
Photon ionisation can be used to produce almost exclusively \HCNp\ from \ce{HCN}, contrary 
to $\text{e}^-$ bombardment \cite{Wisthaler2000} (see also section~\ref{sec:HCNpHNCp}).

The \CNm\ anion has been extensively studied spectroscopically in the IR\cite{Forney1992,Dahlmann2022} 
(vibration) and in the mm-wave range \cite{Gottlieb2007,Amano2008} (rotation) and consequently detected in 
space\cite{Agundez2010}. The \CNm\ anion does not react with \ce{H2} and only forms a weakly bound complex 
at low temperatures\cite{Dahlmann2022}, therefore we assume it only plays a marginal role in cyano-\ce{H2} 
chemistry relevant in this context.   

Due to the simplicity of the \HCN/\HNC\ isomerisation process and the small size of the system it is extensively 
studied theoretically, often employing very high levels of theory or to benchmark new methods  
\cite{Mourik2001,Nguyen2015,Khalouf2019}.

The present astrochemical models have difficulties to reproduce the observed \HCNHp\ /\HCN\ ratios due to 
potentially missing important pathways or key reactions whose reaction rates are not well constrained \cite{Fontani2021}.
In their recent study, \citet{Fontani2021} conclude that in order to get the correct molecular abundances, 
laboratory measurements of reactions (\ref{eq:HCNpH2}), (\ref{eq:HNCpH2}) and (\ref{eq:HCOreact}) are needed. 
This paper focuses on the experimental determination of the reaction rate coefficients for reactions 
(\ref{eq:HCNpH2}), (\ref{eq:HNCpH2}) and (\ref{eq:CNreact}) in the temperature range of 
17 -- 250~K relevant for a variety of astrochemical environments.            

\section{Methods}

\subsection{Experimental setup}

The experimental setup is described in detail in \cite{Jusko2023}, only a short overview will be given here. 
The ions are produced in the storage ion source (SIS) \cite{Gerlich1992}, mass selected by passing through the
first quadrupole mass filter and then refocused on the trap axis using an electrostatic bender. 
The 22 pole radio frequency ion trap is positioned on a cold head (RDK-101E, Sumitomo) enabling operation down to $4\;\rm{K}$. The pressure 
is measured by a Bayard-Alpert type ionisation gauge calibrated by a capacitance manometer CMR 375 from Pfeiffer. After a set 
storage time, the trap is opened and the ions exiting the trap are mass selected by the second quadrupole mass 
spectrometer and detected in a Daly type conversion detector. 
The trapped ions are cooled in collisions with helium gas that is introduced for a few milliseconds into the trap volume 
at the beginning of every trapping cycle by a custom piezo valve. 
This measurement sequence is repeated, usually at $1\;\text{Hz}$
repetition rate, until an adequate signal for the set experimental condition is acquired.

To form \HCNp~and \HNCp~ions, either acetonitrile (\ce{CH3CN}) or vapors of bromium cyanide (\ce{BrCN}) with water,
producing \ce{HCN}, were continuously leaked into the SIS. 
When acetonitrile was used, a portion (about $15\;\%$) of ions with mass $27\;m/z$ did not react with hydrogen and oxygen 
molecules. Here and in the following text $m/z$ denotes mass to charge ratio. We assume that this non-reacting fraction consists of \ce{C2H3+} ions. The production of non reactive species 
was not observed when bromium cyanide was used as a precursor gas in the SIS.

In experiments focused on the reaction of \CNp~with \ce{H2}, the use of acetonitrile led to high fraction of ions with mass $26\;m/z$ 
(almost $50\;\%$) not reacting with \ce{H2} nor \ce{O2}. We tentatively identified these ions as \ce{C2H2+} with reported 
reaction rates with \ce{H2} at least 3 orders of magnitude lower\cite{Smith1993} than for \CNp\ + \ce{H2} and described as
``no reaction/ slow''\cite{Scott2000} for $\ce{C2H2+} + \ce{O2}$, in comparison 
to reaction of \CNp\ with atomic \ce{O} with the reaction rate of $2\times10^{-10}\;\rateU$ \cite{Viggiano1980, Scott2000}.  

\subsection{Ratio of \HCNp\ to \HNCp\ in the ion trap}
\label{sec:HCNpHNCp}

We estimate the relative amount of \HCNp\ to \HNCp\ in the trap using isomer sensitive reaction probing.
\HNCp, the lower energy isomer by $0.94\;\evU$, and \HCNp\ have different reaction channels in ion-molecule reactions with 
\ce{O2}
\begin{equation}
	\HCNp + \ce{O2} \to \ce{O2+} + \HCN,
	\label{eq:HCNO21}
\end{equation}
\begin{subequations}
	\begin{align}
		\HNCp + \ce{O2} &\to \ce{HNCO+} + \ce{O}   \label{eq:HNCO2a}   \\ 
		&\to \ce{NO+} + \ce{HCO},   \label{eq:HNCO2b}
	\end{align}
	\label{eq:HNCO21}
\end{subequations}
and \ce{SF6} 
\begin{equation}
	\HCNp + \ce{SF6} \to \ce{SF5+} + \ce{HF} + \ce{CN}  
	\label{eq:HCNSF61}
\end{equation}
\begin{equation}
		\HNCp + \ce{SF6} \to \ce{HNCF+} + \ce{SF5},    
	\label{eq:HNCSF61}
\end{equation}
with all the species in their vibrational ground states.
The rate coefficients for reactions (\ref{eq:HCNO21}) and (\ref{eq:HNCO21}) were reported at $300\;\rm{K}$ as 
$5.0\times10^{-10}\;\rateU$ and $3.6\times10^{-10}\;\rateU$, respectively \cite{Petrie1991}. 
Reactions with \ce{SF6} (\ref{eq:HCNSF61}) and (\ref{eq:HNCSF61}) were reported at 300~K as $1.3\times10^{-9}\;\rateU$ 
and $1.2\times10^{-9}\;\rateU$, respectively \cite{Petrie1990}. All those reactions are exothermic.

\begin{figure}[h]
\centering
\includegraphics[]{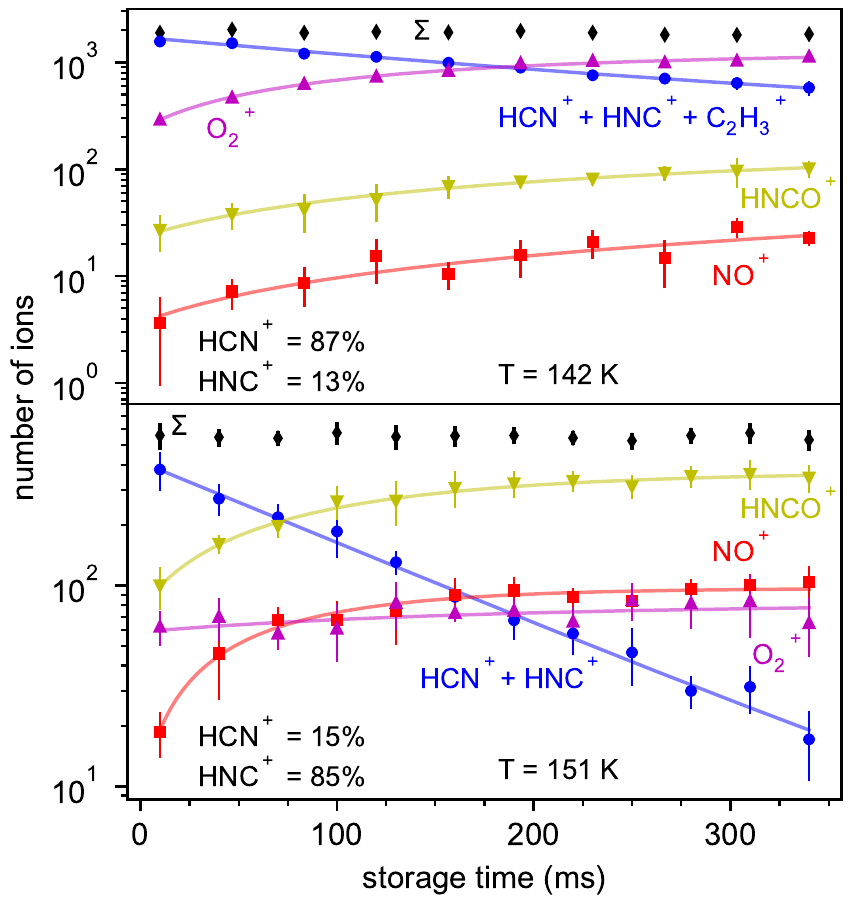}
\caption{\label{fig:IsomersO} \HCNp/\HNCp\ isomer ratio determination using \ce{O_2} probing gas.
Upper panel: conditions with prevalent higher energy \HCNp\ isomer, formed predominantly in the ion source. 
Lower panel: conditions with prevalent \HNCp\ isomer, using the \ce{CO2} conversion reaction~(\ref{eq:HNCCOO}). 
The sum of all the ions in the trap is denoted by $\Sigma$. The full lines are the results of fit of the data 
by solving set of corresponding balance equations. See reactions~(\ref{eq:HCNO21}) and (\ref{eq:HNCO21}) and text.}
\end{figure}

The number of trapped ions as a function of the storage time with a small amount 
of \ce{O_2} gas (ca. $5\times10^{9}\;\text{cm}^{-3}$)
%
%
present in the trap volume is plotted in both panels of Fig. \ref{fig:IsomersO}. The primary ions with mass $27\;m/z$ 
(\HCNp, \HNCp, and \ce{C2H3+}) were produced in the SIS from acetonitrile and then trapped and cooled (translation and vibration)
using the initial He pulse.
As can be seen from the upper panel of 
Fig. \ref{fig:IsomersO}, the main product under these conditions are \ce{O_2+} ions, indicating that the majority of ions with mass 
$27\;m/z$ are \HCNp. 

At low energies, $<0.1\;\evU$, the collision of \HCNp~with \ce{CO} or \ce{CO_2} leads to the formation of the more stable 
\HNCp\ isomer as a result of double charge transfer in the collisional complex \cite{Hansel1998}. 
\ce{CO2} is added into the trap volume to enhance the fraction of the \ce{HNC+} 
isomer inside the trap using the catalytic reaction
\begin{equation}
		\HCNp + \ce{CO2} \to \ce{HNC+} + \ce{CO2} \quad \Delta H = -0.94\;\evU.    
	\label{eq:HNCCOO}
\end{equation}
The number density of \ce{CO_2} used in the experiments (order of magnitude or more than other reactants) 
ensures that the isomerisation reaction is dominant over reactions with \ce{O2} for \HNCp/\HCNp\ ratio estimation or 
with \ce{H2} for reaction rates.

The lower panel of Fig. \ref{fig:IsomersO} illustrates the typical \HNCp\ enrichment using the \ce{CO2} technique.
The dominant product in the \ce{O2} probing reaction is \ce{HNCO+}, implying that the most prevalent isomer is \HNCp.
The actual fractions of \HCNp~and \HNCp~were calculated by solving a set of corresponding balance equations (see section Data analysis) and 
extrapolating the numbers of ions of the given species in the trap to the time of 500 ms.      
Analogous results were obtained with \ce{SF_6} used as a probe gas. 

It is important to note, that the addition of \ce{CO_2} directly to the ion source did not lead to a substantial change in 
the measured fractional populations of \HCNp\ and \HNCp\ in the trap. 
We attribute this behaviour to the relatively high energy of ions in the source. 
Ions just produced by electron bombardment are substantially more energetic than few hundred
meV, which is required for reaction~(\ref{eq:HNCCOO}) to be dominant over a simple no-isomerisation collision.

\vspace{0.5cm}

\subsection{Data analysis}

By integrating the chemical rate equations for the ion number densities over the trap volume, balance equations 
describing time evolutions of the number of trapped ions $n_\mathrm{i}$ are obtained. As an example, for the 
reaction of \HCNp~with \ce{H_2} the corresponding balance equation can be written as
\begin{equation}
	\frac{\mathrm{d}n_\mathrm{HCN^+}}{\mathrm{d}t} = -k^\mathrm{H_2}_\mathrm{HCN^+}[\ce{H_2}]n_\mathrm{HCN^+},    
	\label{eq:balanceeq}
\end{equation} 
where $k^\mathrm{H_2}_\mathrm{HCN^+}$ is the binary rate coefficient for the reaction of \HCNp~with \ce{H_2} 
and $[\ce{H_2}]$ is the \ce{H_2} number density. 
When evaluating the time dependencies of the number of ions in the trap, the quantities that are obtained by 
fitting the set of appropriate balance equations to the measured data are reaction rates, \ie, the reaction rate 
for reaction~(\ref{eq:balanceeq}) is $k^\mathrm{H_2}_\mathrm{HCN^+}[\ce{H_2}]$. The reaction rate coefficient 
is then determined from the slope of the dependence of the reaction rate on the number density of the 
corresponding reactant at the given temperature (for a detailed description of the fitting procedure see 
ref.\cite{Jusko2023}).
Throughout the text, quoted uncertainties are statistical errors of the corresponding fitting procedures. 
The systematic error, arising mainly from the uncertainty in pressure calibration, is estimated to be 20~\% \cite{Jusko2023}.  

\section{Results and discussions}
\subsection{\HCNp~+~\ce{H_2}}

\begin{figure}[h]
	\centering
	\includegraphics[]{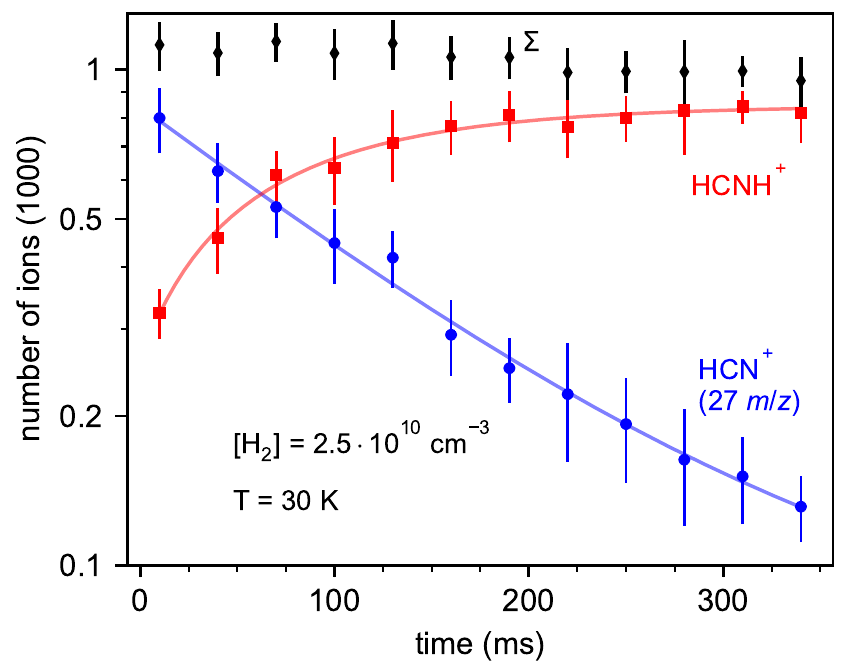}
	\caption{\label{fig:HCNpnumbers}
		Number of ions as a function of trapping time. Typical number of \HCNp\ reactants (contains traces 
		of \HNCp\ and \ce{C2H3+}) and \ce{HCNH^+} product ions shown 
		at the trap temperature of $T = 30\;\text{K}$. 
		Deviation from ideal exponential decay to exponential decay with offset,
		caused by the non reactive \ce{C2H3+}, can be seen around $300\;\text{ms}$.
		The sum of all stored ions is denoted by $\Sigma$. 
	} 
\end{figure}

\begin{figure}[h]
	\centering
	\includegraphics[]{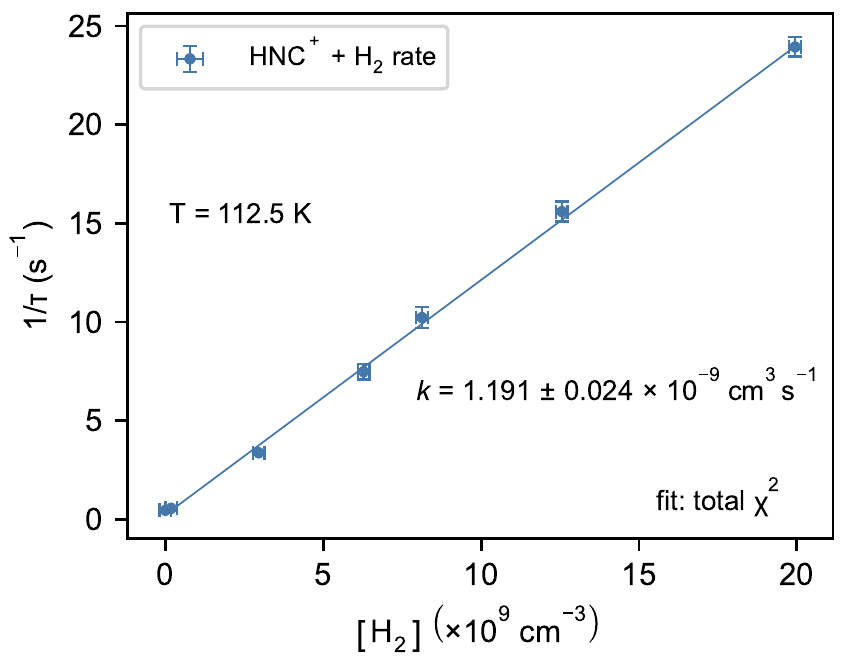}
	\caption{\label{fig:HNCnumDen}
		Typical rate of the reaction as a function of \ce{H2} number density in the trap, here depicted for reaction (\ref{eq:HNCpH2})
		at the trap temperature of $T = 112.5\;\text{K}$. The slope corresponds to the rate coefficient $k$ 
		(see text and ref.\cite{Jusko2023}).
	} 
\end{figure}

The reaction of \HCNp\ with \ce{H_2} was studied in the temperature range of $17 - 250\;\text{K}$. An example of the measured 
time dependence of the number of ions in trap when \HCNp\ was the dominant isomer is shown in Fig.~\ref{fig:HCNpnumbers}. 
As the \HCNp\ ions react with molecular hydrogen, \ce{HCNH^+} ions are formed as the only product of the reaction.
Similar data were obtained at each temperature for several values of hydrogen number density in order to evaluate
the reaction rate coefficients (see Fig.~{\ref{fig:HNCnumDen} for a typical reaction rate plot as a function of number density}). 
The results are plotted in Fig.~\ref{fig:k} as down facing triangles. The collisional
Langevin reaction rate coefficient for this reaction is $1.54\times 10^{-9}~\rateU$ using \ce{H_2} polarisability 
from ref\cite{Milenko1972}. As can be seen from Fig.~\ref{fig:k}, the obtained value of the reaction rate coefficients
are constant in the studied temperature range and slightly lower than the Langevin reaction rate coefficient. 

The reaction of \HCNp\ with \ce{H_2} was previously studied by \citet{Petrie1991} at $300\;\text{K}$ using a selected-ion flow tube 
(SIFT) experiment. \ce{CF_4} and \ce{SF_6} were used as probe gases to distinguish between 
\HCNp\ and \HNCp\ ions in the SIFT experiments \cite{Petrie1990,Petrie1991}. Contrary to what \citet{Petrie1990} thought 
at that time, the ground vibrational state of \HCNp\ does
not react with \ce{CF4} \cite{Hansel1998} indicating that a substantial fraction of \HCNp\ ions in their experiment 
\cite{Petrie1991} was, in fact, vibrationally excited. This could be a possible explanation for the almost $40\;\%$ difference 
between the SIFT data and the reaction rate coefficient obtained at the highest temperature of $250\;\text{K}$ in the present study.     

\subsection{\HNCp\ + \ce{H_2}}
\label{sec:HNCp}

In order to study the reaction of \HNCp\ ions with \ce{H_2}, we first trapped the \HCNp\ ions (with a small fraction of 
non reactive \ce{C_2H_3^+} ions) and converted them to \HNCp\ ions inside the 22 pole trap by applying the 
isomerization reaction~(\ref{eq:HNCCOO}) with \ce{CO2}\cite{Hansel1998}. The \ce{CO2} number density was high 
enough to ensure that the majority of the \HCNp\ ions was converted to \HNCp\ before the measurement of the reaction with 
\ce{H2} was performed. 
The only observed product of the reaction were \ce{HCNH^+} ions.  
For temperatures above 80~K, the \ce{CO_2} gas was added directly into the trap, as the vapour pressure of 
\ce{CO_2} was sufficient ($4\times10^{-6}$ Pa at 80~K\cite{Bryson1974}). 
For lower temperatures, 30~K and 70~K, the \ce{CO_2} had to be mixed with the helium buffer gas and pulsed at the 
beginning of each trapping cycle. 
The charge transfer reaction between \ce{N2^+} and \ce{CO_2} was used to check and confirm the presence of 
sufficient \ce{CO2} number densities. While no depletion of \ce{CO2} was observed down to 70~K, 
below this temperature the \ce{CO2} number density started to decrease reaching our detection limit between $30-40\;\text{K}$.  
Although the value of the reaction rate coefficient obtained for temperature of 30~K corresponds to a mixture of ions 
with mass $27\;m/z$ dominated by \HNCp, we are unable to quantify the isomeric ratio.

\begin{figure}[h]
	\centering
	\includegraphics[]{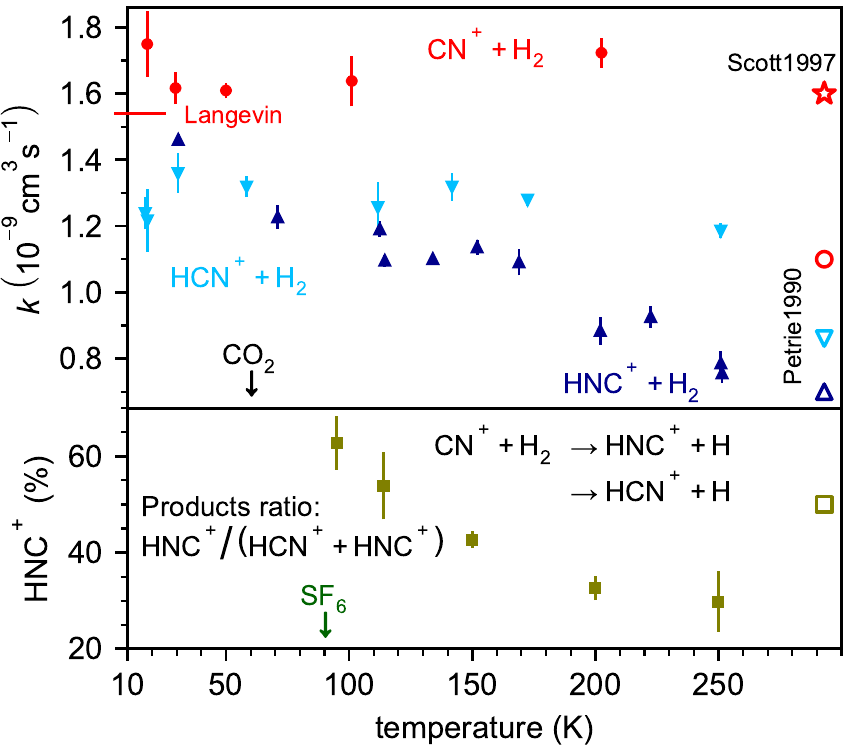}
	\caption{\label{fig:k}
		Reaction rate coefficients for the reactions of \CNp\ (full red circles), \HCNp\ (full cyan down facing triangles) and 
		\HNCp\ (full dark blue up facing triangles) with \ce{H2} (top) and 
		branching ratio for the $\CNp+\ce{H2}$ reaction (bottom). Open symbols denote values obtained in previous studies 
		(star \cite{Scott1997}, the rest \cite{Petrie1991}). The corresponding value of the Langevin reaction rate 
		coefficient is $1.54\times 10^{-9}~\rateU$ (for all the reactions). 
		The down pointing arrows labelled \ce{CO2} and \ce{SF6} show the freeze 
		out temperature of the respective reactants (for details see text). 
	} 
\end{figure}

The measured value of the reaction rate coefficient is approximately half of the collisional Langevin rate coefficient 
at 250~K and increases with decreasing temperature (see Fig.~\ref{fig:k}). The 300~K SIFT value 
reported by \citet{Petrie1990} is in very good agreement with our data.

\subsection{\CNp\ + \ce{H_2}}            

The \CNp\ ions were produced in the SIS from bromium cyanide with an admixture of water vapours. The majority of the trapped ions
with mass $26\;m/z$ reacted with hydrogen, which was leaked directly into the trap, implying that there was only a small fraction
of \ce{C_2H_2^+} ions present in the trap during these experiments \cite{Hansel1998}. The value of the reaction rate coefficient
for reaction of \CNp\ with \ce{H_2} was measured in the temperature range of $17 - 200\;\text{K}$. The results are shown in 
Fig.~\ref{fig:k}. The obtained value is close to the collisional Langevin reaction rate coefficient. No significant temperature 
dependence was observed. Our data are in excellent agreement with the study by \citet{Scott1997} performed at 300~K.

The reaction of \CNp\ with \ce{H_2} results in the production of \HCNp\ and \HNCp\ ions. As both these products subsequently react with 
hydrogen (see above), it was impossible to determine the product branching ratio with hydrogen continuously added into the trap.
Instead, we added a small amount of \ce{H_2} (approximately $0.2\;\%$) to the short helium pulse that is used to cool down ions at 
the beginning of each trapping period. In this way, we were able to maximize the amount of produced primary ions (\HCNp\ and 
\HNCp) and to minimize the subsequent formation of secondary ions (\ce{HCNH^+}). 

\begin{figure}[h]
	\centering
	\includegraphics[]{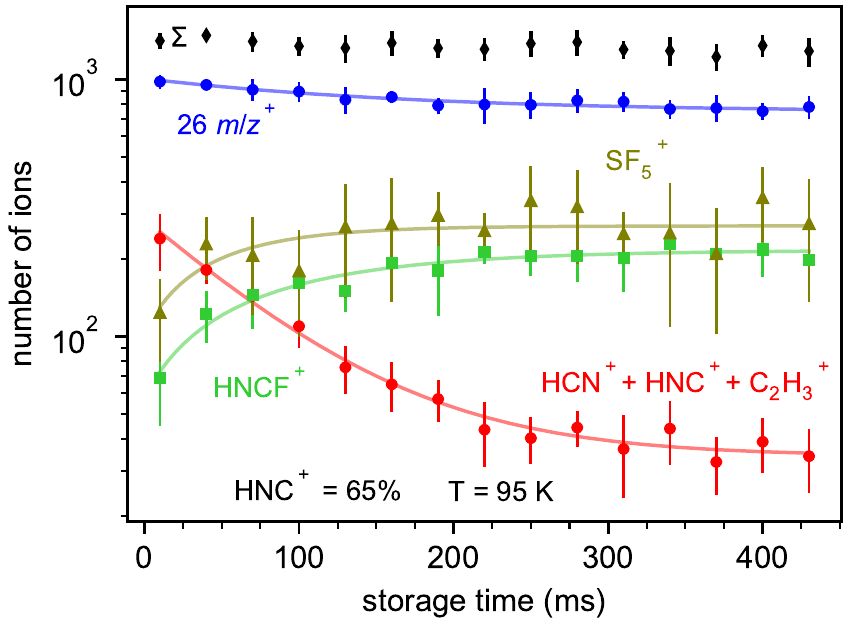}
	\caption{\label{fig:ratio} Determination of the ratio of the products of the $\CNp+\ce{H2}$ reaction.
		\ce{SF6} was used as a reaction product isomer probing gas. 
		The \ce{SF5+} ions are products of the reaction of either \CNp\ or \HCNp\ with \ce{SF6}.
		Only ground state \HNCp\ produces \ce{HNCF+} in reaction with \ce{SF6}. 
		The \HNCp\ fraction is determined from the increase of number of \HNCFp\ ions compared to the decrease of number 
		of ions with mass $27\;m/z$. 
		The primary mass $26\;m/z$ corresponds to an undetermined ratio of \CNp\ and \ce{C2H2+}; the latter is non reactive.
		The ion signal for \ce{SF5+} is multiplied by a factor of 5.5 to account for the 
		sensitivity of the detection system.
	} 
\end{figure}

The actual product branching ratio was determined by utilizing different reactivity of \HCNp\ and \HNCp\ ions with \ce{SF_6},
which was added directly into the trap. An example of the measured number of ions in the trap as a function of storage time is 
shown in Fig.~\ref{fig:ratio}. \ce{SF5+} is formed in reactions of \CNp\ as well as \HCNp\ and vibrationally excited \HNCp\ 
with \ce{SF_6}, \ie, this channel is not suitable for product isomer probing.
Solely \HNCp\ ions in the ground state reacting with \ce{SF6} do form \ce{HNCF+}. Therefore, the \HNCp\ fraction was 
determined from the increase of the number of \ce{HNCF+} ions compared to the decrease of the ions of mass $27\;m/z$. 
The measured \HNCp/\HCNp\ product branching ratios are shown in lower panel of Fig.~\ref{fig:k}.
At 250~K, almost $70\;\%$ of all ions produced in reaction of \CNp\ with \ce{H_2} are \HCNp. As the temperature decreases, 
reaction~(\ref{eq:CNreact}) results in larger fraction of produced \HNCp\ ions and below 100~K the \HNCp\ channel of 
reaction~(\ref{eq:CNreact}) accounts for more than $60\;\%$ of total produced ions. 

This is in disagreement with results of \citet{Petrie1990} who observed the same probability of both channels of reaction 
(\ref{eq:CNreact}) at 300~K.
As discussed above in relation with the study of the reaction of \HCNp\ with \ce{H2}, it is possible that the \CNp\ ions in
the study by \citet{Petrie1990} were vibrationally excited. While we are absolutely certain that in our experiment the primary 
\CNp\ ions are in their vibrational ground state, we can not rule out that some of the \HCNp\ and \HNCp\ ions produced inside 
the trap (reaction~(\ref{eq:CNreact})) possess some vibrational excitation. As reaction (\ref{eq:CNreactHNC}) is exothermic 
by $2.28\;\eV$ \cite{Scott1997}, \HNCp\ ions can have enough internal energy available to form \ce{SF5+} ions in reaction with 
\ce{SF6} \cite{Hansel1998} and thus influence our data analysis to overestimate the \HCNp\ fraction of produced ions.
Therefore, our \HNCp/\HCNp\ product branching ratios shall be interpreted as lower limit; the ratio may be higher, but not 
lower than reported.

\subsection{\CNp\ + \ce{O_2}}

The reaction rate coefficient for reaction
\begin{subequations}
	\begin{align}
		\CNp + \ce{O2} &\to \ce{O2+} + \ce{CN}   \label{eq:CNO2a}   \\
		&\to \ce{NCO+} + \ce{O}   \label{eq:CNO2b}   \\ 
		&\to \ce{NO+} + \ce{CO}   \label{eq:CNO2c}
	\end{align}
	\label{eq:CNO21}
\end{subequations}
was determined between 100 and 230~K. Acetonitrile (all temperatures) and bromium cyanide with admixture of water vapours (152~K)
were used as precursors to form \CNp\ ions in the SIS (see Fig.~\ref{fig:other_ks}). 
The major product of the reaction is \ce{O2+}, accounting for more than $80\;\%$ of the produced ions, followed by \ce{NCO+}
(less than $20\;\%$ of product ions) and \ce{NO+} (few percent of produced ions). \ce{NCO+} ions reacted slowly with \ce{O2},
complicating the determination of the product branching ratios. For comparison, \citet{Raksit1984} reported the product 
branching ratios for reaction (\ref{eq:CNO21}) as 0.6:0.2:0.2 at room temperature.

The measured rate coefficient shows a very steep increase with decreasing temperature and is in a good agreement with the previous 300~K
SIFT experimental value \cite{Raksit1984}.
Although the corresponding collisional Langevin reaction rate coefficient $k_\mathrm{L}=7.8\times 10^{-10}~\rateU$ is 
of the same order of magnitude as the experimentally determined rate coefficient, 
the comparison is only made for reference as the measured dependence 
implies a barrier in the reaction path of the overall exothermic reaction.  

\subsection{Attempt at electronic spectroscopy of \HNCp}

We attempted to perform electronic spectroscopy of \HNCp\ cation in photon energy range $1.6 - 2.5\;\eV$
using a laser induced charge transfer (LICT) action scheme of ions stored in the 22 pole trap at 
ca. $150\;\text{K}$ to avoid any kind of neutral freeze out. The range has been selected 
for $\rm{X}\,^{2}\Sigma^{+} \to \rm{A}\, ^{2}\Pi$ \HNCp\ transition predicted 
by previous computational results around $2\;\eV$ \cite{Gans2019}.
LICT of \HNCp\ to \ce{Xe} ($\Delta\sim 0.1\;\eV$) as well as to \ce{CO2} ($\Delta\sim 1.6\;\eV$, note that \ce{CO2} is more
suitable for LICT studies of higher energy isomer \HCNp)\cite{Hansel1998} 
has been attempted using a supercontinuum laser\cite{Jusko2023} with \HNCp\ ions produced as described in section~\ref{sec:HNCp}.
Unfortunately no action spectroscopic signal could be recorded.
While the aforementioned LICT schemes for IR vibrational and VIS electronic studies of \HNCp\ should be straightforward and 
very effective, the lack of experimental spectroscopic data for \HCNp/\HNCp\ in gas phase remains remarkable.

\subsection{Comparison to other processes involved in \HCNHp\ formation in the ISM}

\begin{figure}[]
\centering
\includegraphics[]{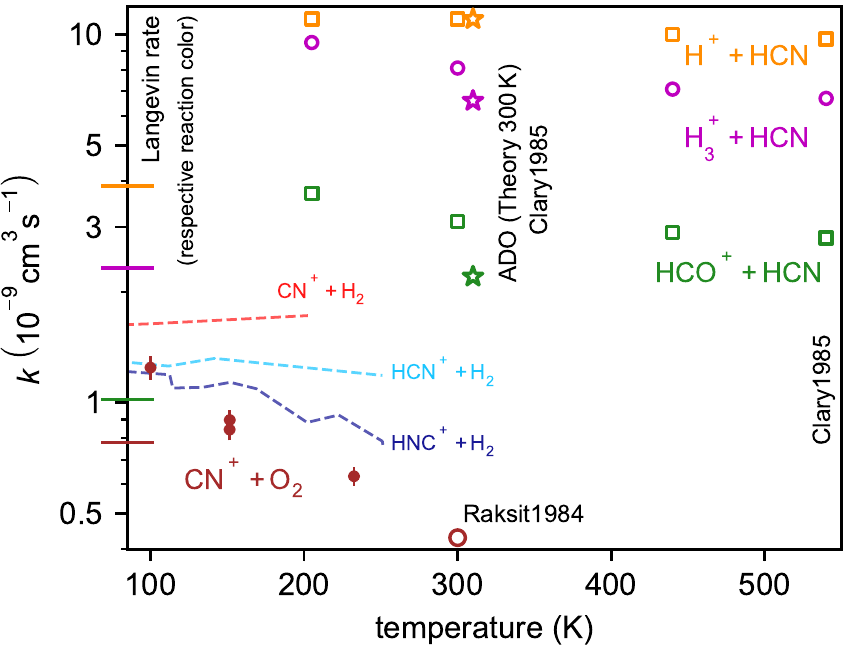}
\caption{\label{fig:other_ks}
  Rate coefficients of ion-molecule reactions leading to \HCNHp\ as a function of temperature.
  Data for reaction of \CNp, \HCNp, and \HNCp\ with \ce{H2} from Fig.~\ref{fig:k} (dashed lines).
  Previous experimental data for \ce{H+}, \ce{H3+}, and \ce{HCO+} with \ce{HCN} (orange, magenta, green open symbols) \cite{Clary1985}, 
  including the theoretical average dipole orientation (ADO) approximation (stars)  \cite{Clary1985}.
  Reaction rate of $\CNp + \ce{O2}$ (brown), with previous room temperature SIFT value (brown open circle) \cite{Raksit1984}
  shown for reference. Langevin reaction rates are illustrated by horizontal lines on the left abscissa.
  }
\end{figure}

A comparison of the reaction rate coefficients obtained in the present study with those measured for other \HCNHp\ formation 
processes by \citet{Clary1985} is shown in Fig. \ref{fig:other_ks}. In cold, dense regions of the interstellar medium, the most important gas phase processes for the formation of \HCNp\ and \HNCp\ ions are considered to be \cite{Quenard2017} 
reaction (\ref{eq:CNreact}) and the charge transfer reaction
\begin{equation}
	\ce{H^+} + \ce{HCN}/ \ce{HNC} \to \HCNp/ \HNCp + \ce{H}.  
	\label{eq:HpHCN}
\end{equation}

\citet{Clary1985} studied reaction (\ref{eq:HpHCN}) for HCN reactant and reported values of the reaction rate coefficient 
close to $1\times10^{-8}~\rateU$ and practically independent on temperature between 205 and 540~K. The calculated exothermicity 
of reaction (\ref{eq:HpHCN}) is $8 \pm 10\;\text{meV}$ \cite{Clary1985}, therefore reaction (\ref{eq:HpHCN}) could be slightly 
endothermic and the value of its reaction rate coefficient at temperatures close to 10~K much lower than that reported 
by \citet{Clary1985} at higher temperatures. In that case, reaction (\ref{eq:CNreact}) proceeding with almost Langevin reaction rate coefficient, 
would be the dominant process for \HCNp\ formation in such environments.      

Astrochemical databases such as KIDA \cite{Wakelam2012} usually contain the 
reaction rate coefficients for reactions (\ref{eq:HCNpH2}), (\ref{eq:HNCpH2}) and (\ref{eq:CNreact}) that were measured 
by \citet{Petrie1991} at 300~K with 
an unknown fraction of vibrationally excited ions. According to the present data obtained with vibrationally cold ions 
at low temperatures, the actual values of reaction rate coefficients for reactions (\ref{eq:HCNpH2}) and (\ref{eq:CNreact}) 
are higher by a factor of up to 1.6. In the case of low temperature \HCNHp\ formation in the reaction of \HNCp\ 
with \ce{H_2} (\ref{eq:HNCpH2}) the ratio between our value of the reaction rate coefficient and that in the 
KIDA entry is almost two. As a result, the astrochemical models employing these database values are probably underestimating 
the \HCNp/\HNCp\ and \HCNHp\ formation in cold, dense regions of the interstellar medium.

The branching ratio for reaction (\ref{eq:CNreact}) is 1:1 in the KIDA database in accordance with the 300~K value 
reported by \citet{Petrie1991}. Although our data show that at 250~K almost 70~\% of products of reaction (\ref{eq:CNreact}) are
\HCNp\ ions, the branching ratio is close to that in 
the KIDA database around 100~K. Unfortunately, we were not able to measure the branching ratio of 
reaction (\ref{eq:CNreact}) to lower temperatures due to the employed chemical probing scheme. 
Given the observed temperature dependence, it is possible that in regions of cold interstellar gas, 
where reaction (\ref{eq:CNreact}) is a key formation process for \HCNp\ and \HNCp\ ions, the actual branching ratio strongly 
favours \HNCp\ production.

\subsection{Parameterized reaction rates coefficients}

The temperature dependent reaction rate coefficients of reactions 
(\ref{eq:HCNpH2}), (\ref{eq:HNCpH2}), (\ref{eq:CNreact}), and, (\ref{eq:CNO21}) measured in this work have been parameterized
using the Arrhenius–Kooij formula as defined in equation~(1) in ref. \cite{Wakelam2012}. The results of the least square fit 
together with the temperature range where the fit is valid, are reported in Table~\ref{t:kida}. 
The $\CNp + \ce{O2}$ room temperature measurement \cite{Raksit1984} has been included in the fit.
Detailed procedures are available in the data set associated with this work (see section Data Availability). 

\begin{table}[!h]
    \caption[]{Parameterized reaction rate coefficients using Arrhenius–Kooij formula\cite{Wakelam2012} $k(T) = A(T/300)^B \exp(-C/T)$.
    }\label{t:kida}
    \begin{center}
    \sisetup{separate-uncertainty}
    \begin{tabular}{lclS[table-format = 3.2(3)]S[table-format = 4.3(3)]S[table-format = 2.0(2)]}
        \toprule
        Reaction        &  ~~~~~~~~~~~~      & $T_{\text{range}}$  (K)   &  $A^a$           &           $B$        &        $C^c$     \\
        \midrule
        $\CNp+\ce{H2}$  & (\ref{eq:CNreact}) & $10 - 300$                &  16.67\pm0.29    &                      &                  \\
        \noalign{\smallskip}
        $\CNp+\ce{O2}$  & (\ref{eq:CNO21})   & $95 - 300$                &  4.72\pm0.26     & -0.889\pm0.075       &                  \\
        \noalign{\smallskip}
        $\HCNp+\ce{H2}$ & (\ref{eq:HCNpH2})  & $10 - 300$                &  12.7\pm0.02     &                      &                  \\
        \noalign{\smallskip}
        $\HNCp+\ce{H2}$ & (\ref{eq:HNCpH2})  & $10 - 90$                 &  12.7\pm0.02$^\&$  &                      &                  \\
                        &                    & $90 - 300$                &  9.52\pm0.69     &    -1.02\pm0.27      &   88\pm34        \\
        \bottomrule
    \end{tabular}
    \vskip 0.2em
    \end{center}
    {\footnotesize{\emph{Note:}
    $a$ -- in $10^{-10}\;\text{cm}^3\,\text{s}^{-1}$; $c$ -- in $\text{K}$; $\&$ -- same as the reaction $\HCNp+\ce{H2}$;
    The statistical errors of the least square fit are reported.
    }}
\end{table}

\normalsize

\clearpage
\newpage
\section{Conclusion}
\vspace{0.5cm}

The formation and destruction of \HCNp\ and \HNCp\ isomers in reactions with \ce{H_2} were studied in the temperature 
range of $17 - 250\;\text{K}$. The values of the reaction rate coefficients for reactions of \CNp\ and \HCNp\ with \ce{H_2} are 
constant in the studied temperature range and close to the collisional Langevin reaction rate coefficient. The reaction 
of \CNp\ with \ce{H2} produces predominantly \HCNp\ at 250~K, but below 100~K, \HNCp\ is favoured. The obtained 
values of the reaction rate coefficient for the reaction of \HNCp\ with \ce{H_2} decrease with increasing temperature.
As current astrophysical databases such as KIDA \cite{Wakelam2012} contain rate coefficients for these key reactions that were obtained at 300 K, in some cases with vibrationally excited ions, we believe that our results will help to improve models of cyanide chemistry in the interstellar medium. 

The isomer specific results in this work were achieved solely using chemical probing. This technique is fully dependent
on the availability of chemicals with the right energy levels with respect to the isomers being studied. Even then, 
the process is rather tedious, since several reactants have to be used in sequence (\eg\ isomerisation/ reactivity/ 
probing) with tight control of the real number density inside the trap, especially while lowering the temperature 
towards the neutral gas freeze out. While we have demonstrated that rf multipole ion traps are a suitable tool for 
isomer specific studies, the application of novel techniques and schemes based on direct discrimination of ions with different 
kinetic energy (\eg\ the reaction of of \CNp\ with \ce{H2} releases two equal $m/z$ products with two distinctive total energies), 
as described in \citet{Jusko2023},
or indirectly, based on the transfer of the internal excitation to translation energy in a collision with a neutral, as described 
in the ``leak out'' method \cite{Schmid2022}, shall be developed.

\begin{acknowledgments}

This work was supported by the Max Planck Society, Czech Science Foundation (GACR 22-05935S) and COST Action CA17113.
The authors gratefully acknowledge the work of the electrical and mechanical workshops and engineering 
departments of the Max Planck Institute for Extraterrestrial Physics.

\end{acknowledgments}

\section*{AUTHOR DECLARATIONS}

\subsection*{Conflict of Interest}
The authors have no conflicts to disclose.

\subsection*{Author Contributions}
Paola Caselli: Conceptualization, Funding acquisition, Investigation, Writing -- review \& editing.
Petr Dohnal: Conceptualization, Funding acquisition, Investigation, Writing -- original draft.
Miguel Jim{\'e}nez-Redondo: Conceptualization, Data curation, Investigation, Writing -- review \& editing.
Pavol Jusko: Conceptualization, Data curation, Investigation, Visualization, Writing -- original draft.

\section*{DATA AVAILABILITY}
The data that support the findings of this study are openly available at \\
\url{http://doi.org/10.5281/zenodo.7704359}.


%
%

\end{document}